\documentclass[twoside,twocolumn,9pt]{article}
\usepackage{extsizes}
\usepackage[super,sort&compress,comma]{natbib} 
\usepackage[version=3]{mhchem}
\usepackage[left=1.5cm, right=1.5cm, top=1.785cm, bottom=2.0cm]{geometry}
\usepackage{balance}
\usepackage{mathptmx}
\usepackage{sectsty}
\usepackage{graphicx} 
\usepackage{lastpage}
\usepackage[format=plain,justification=justified,singlelinecheck=false,font={stretch=1.125,small,sf},labelfont=bf,labelsep=space]{caption}
\usepackage{float}
\usepackage{fancyhdr}
\usepackage{fnpos}
\usepackage[english]{babel}
\addto{\captionsenglish}{%
  
}
\usepackage{array}
\usepackage{droidsans}
\usepackage{charter}
\usepackage[T1]{fontenc}
\usepackage[usenames,dvipsnames]{xcolor}
\usepackage{setspace}
\usepackage[compact]{titlesec}
\usepackage{hyperref}
\usepackage[most]{tcolorbox}
\usepackage{bbold}

\usepackage{epstopdf}

\definecolor{cream}{RGB}{222,217,201}
\begin{document}

\pagestyle{fancy}
\thispagestyle{plain}
\fancypagestyle{plain}{
\renewcommand{\headrulewidth}{0pt}
}

\makeFNbottom
\makeatletter
\renewcommand\LARGE{\@setfontsize\LARGE{15pt}{17}}
\renewcommand\Large{\@setfontsize\Large{12pt}{14}}
\renewcommand\large{\@setfontsize\large{10pt}{12}}
\renewcommand\footnotesize{\@setfontsize\footnotesize{7pt}{10}}
\renewcommand\scriptsize{\@setfontsize\scriptsize{7pt}{7}}
\makeatother

\renewcommand{\thefootnote}{\fnsymbol{footnote}}
\renewcommand\footnoterule{\vspace*{1pt}%
\color{cream}\hrule width 3.5in height 0.4pt \color{black} \vspace*{5pt}} 
\setcounter{secnumdepth}{5}

\makeatletter 
\renewcommand\@biblabel[1]{#1}            
\renewcommand\@makefntext[1]%
{\noindent\makebox[0pt][r]{\@thefnmark\,}#1}
\makeatother 
\renewcommand{\figurename}{\small{Fig.}~}
\sectionfont{\sffamily\Large}
\subsectionfont{\normalsize}
\subsubsectionfont{\bf}
\setstretch{1.125} 
\setlength{\skip\footins}{0.8cm}
\setlength{\footnotesep}{0.25cm}
\setlength{\jot}{10pt}
\titlespacing*{\section}{0pt}{4pt}{4pt}
\titlespacing*{\subsection}{0pt}{15pt}{1pt}

\fancyfoot{}
\fancyfoot[LO,RE]{\vspace{-7.1pt}\includegraphics[height=9pt]{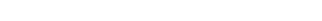}}
\fancyfoot[CO]{\vspace{-7.1pt}\hspace{13.2cm}\includegraphics{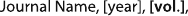}}
\fancyfoot[CE]{\vspace{-7.2pt}\hspace{-14.2cm}\includegraphics{head_foot/RF}}
\fancyfoot[RO]{\footnotesize{\sffamily{1--\pageref{LastPage} ~\textbar  \hspace{2pt}\thepage}}}
\fancyfoot[LE]{\footnotesize{\sffamily{\thepage~\textbar\hspace{3.45cm} 1--\pageref{LastPage}}}}
\fancyhead{}
\renewcommand{\headrulewidth}{0pt} 
\renewcommand{\footrulewidth}{0pt}
\setlength{\arrayrulewidth}{1pt}
\setlength{\columnsep}{6.5mm}
\setlength\bibsep{1pt}

\makeatletter 
\newlength{\figrulesep} 
\setlength{\figrulesep}{0.5\textfloatsep} 

\newcommand{\topfigrule}{\vspace*{-1pt}%
\noindent{\color{cream}\rule[-\figrulesep]{\columnwidth}{1.5pt}} }

\newcommand{\botfigrule}{\vspace*{-2pt}%
\noindent{\color{cream}\rule[\figrulesep]{\columnwidth}{1.5pt}} }

\newcommand{\dblfigrule}{\vspace*{-1pt}%
\noindent{\color{cream}\rule[-\figrulesep]{\textwidth}{1.5pt}} }

\makeatother

\twocolumn[
  \begin{@twocolumnfalse}
{\includegraphics[height=30pt]{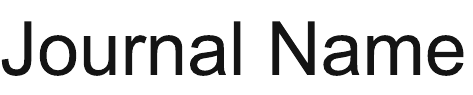}\hfill\raisebox{0pt}[0pt][0pt]{\includegraphics[height=55pt]{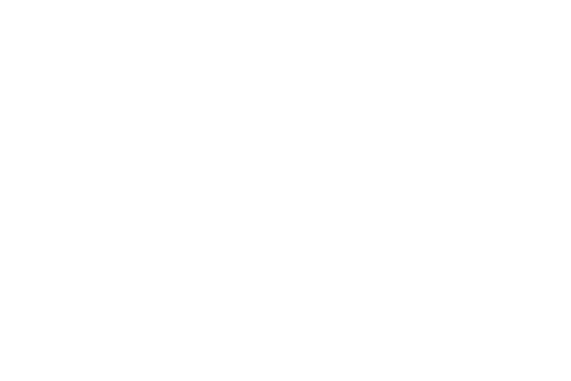}}\\[1ex]
\includegraphics[width=18.5cm]{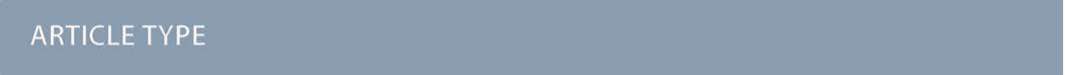}}\par
\vspace{1em}
\sffamily
\begin{tabular}{m{4.5cm} p{13.5cm} }

\includegraphics{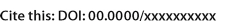} & \noindent\LARGE{\textbf{Impact of light-matter coupling strength on the efficiency of microcavity OLEDs: A unified quantum master equation approach}} \\
 & \vspace{0.3cm} \\

 & \noindent\large{Olli Siltanen,$^{\ast}$\textit{$^{a}$} Kimmo Luoma,\textit{$^{b}$} and Konstantinos S. Daskalakis\textit{$^{a}$}} \\

\includegraphics{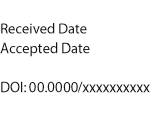} & \\

\end{tabular}

 \end{@twocolumnfalse} \vspace{0.6cm}

  ]

\renewcommand*\rmdefault{bch}\normalfont\upshape
\rmfamily
\section*{}
\vspace{-1cm}


\footnotetext{\textit{$^{a}$~Department of Mechanical and Materials Engineering, University of Turku, Turku, Finland; E-mail: olmisi@utu.fi}}
\footnotetext{\textit{$^{b}$~Department of Physics and Astronomy, University of Turku, Turku, Finland. }}





\sffamily{\textbf{Controlling light-matter interactions is emerging as a powerful strategy to enhance the performance of organic light-emitting diodes (OLEDs). By embedding the emissive layer in planar microcavities or other modified optical environments, excitons can couple to photonic modes, enabling new regimes of device operation. In the weak-coupling regime, the Purcell effect can accelerate radiative decay, while in the strong-coupling regime, excitons and photons hybridize to form entirely new energy eigenstates with altered dynamics. These effects offer potential solutions to key challenges in OLEDs, such as triplet accumulation and efficiency roll-off, yet demonstrations in the strong-coupling case remain sparse and modest. To systematically understand and optimize photodynamics across the different coupling regimes, we develop a unified quantum master equation model for microcavity OLEDs. The model is then applied to estimate device performance in the different coupling regimes to determine which one is the best.}}\\


\rmfamily 


\section*{Introduction}

Organic light-emitting diodes (OLEDs) are revolutionizing display and lighting applications with their unique advantages over traditional inorganic LEDs. Producing vibrant colors, achieving high-contrast ratios, and operating on flexible substrates, OLEDs have become the cornerstone of next-generation devices such as foldable smartphones and transparent displays~\cite{Forrest2004,Li2017,Forrest2020,Huang2020,Zhu2020,Hong2021,Chen2023}. In addition, it can be eco-friendlier to manufacture and recycle OLEDs than LEDs~\cite{Volz2015,Franz2017,Schulte2019,Palo2023}. Despite their benefits, there are some inherent challenges with OLEDs that have hindered their adoption in a wider range of applications, particularly in general illumination. Most notably, excitons---bound electron-hole pairs responsible for light emission---can exist in two fundamentally different spin configurations: one singlet state and three triplet states~\cite{Yersin2011}. The singlet exciton is the only one that can efficiently and rapidly emit light through fluorescence. In contrast, triplet excitons cannot directly emit photons due to spin conservation rules, making them non-emissive in typical fluorescent OLEDs~\cite{Forrest2020}. The non-emitting triplet states not only fail to contribute to light output but also pose additional challenges. At higher input currents and exciton densities, the long-lived triplet states are more likely to interact with other excitons and polarons, annihilate, and reduce device efficiency---a phenomenon known as efficiency roll-off~\cite{Giebink2008,Murawski2013,Diesing2024}. Furthermore, the intermediate encounter complexes can reach energies high enough to break molecular bonds and cause irreversible degradation of the organic materials~\cite{Tankeleviciute2024}.

\begin{figure*}[t!]
\includegraphics[width=\textwidth]{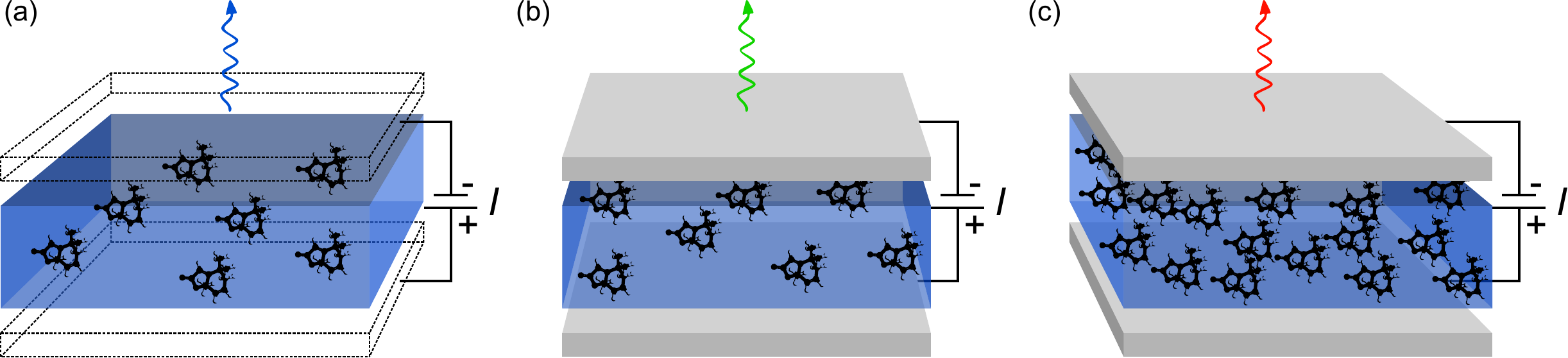}
\caption{\label{fig:schematics} Schematic picture of the study. (a) A basic OLED. The mirrors (or their reflectivity) can be ignored. (b) A weakly coupled microcavity OLED. The singlet excitons emit through optical modes supported by the cavity, with the emission possibly enhanced and red-shifted. (c) A strongly coupled microcavity OLED. The number of molecules is increased (or the mode volume decreased) to enter a regime, where emission through polaritons might be even more enhanced and red-shifted. The figure illustrates the research question: \textit{Does stronger light-matter coupling translate into better device performance?} To answer this, we first need to develop a unified master equation model.
}
\end{figure*}

Traditional molecular design techniques allow to battle the issues outlined above up to some extent. Improving the triplet-to-singlet conversion rate often comes with reduced oscillator strength and therefore reduced emission rate. Moreover, even the most efficient thermally activated delayed fluorescence (TADF) materials experience substantial efficiency roll-off at relatively low injection currents~\cite{Diesing2024}. Alternatively, triplet-to-singlet conversion and emission can be controlled with planar microcavities~\cite{Dirr1998,Xiang2013,Murawski2014,Ye2021,Zhao2024,Hassan2025,Wei2025,Manish2025,Ahmed2025}. By embedding the emitting molecules within optical cavities and engineering their photonic environment, it becomes possible to control exciton dynamics and enhance light emission. In the weak-coupling regime, the Purcell effect increases the radiative decay rate of singlet excitons, thereby improving overall emission efficiency~\cite{Vahala2003}. In the strong-coupling regime, where exciton-photon interactions exceed exciton-photon losses, light and matter hybridize into polaritons---collective light-matter states that open up new avenues for controlling exciton dynamics and boosting device efficiency~\cite{Bhuyan2023,Mischok2023,Abdelmagid2024,Yuan2024,Sandik2024,Siltanen2025}. While polariton physics is an active field, most studies have focused on fundamental phenomena rather than device-level performance (see, e.g., Ref.~\cite{Perez2025}). As a result, the potential of polaritons remains underexplored in practical OLED architectures. Our work addresses this gap.

To determine which coupling regime enhances OLEDs the most, we develop the first quantum master equation model that spans across all of them. While previous models have focused on specific regimes or processes~\cite{Rebentrost2009,Nakano2016,Herrera2017,Takahashi2019,Martinez-Martinez2019,Gu2021,Casanova2022}, our model provides a comprehensive understanding of light-emission mechanisms across different OLEDs architectures. The OLED types considered are summarized in Fig.~\ref{fig:schematics}. Our unified approach not only advances the theoretical foundation of microcavity OLEDs but also provides practical guidelines for optimizing device efficiency. 

\section*{Results and discussion}

\subsection*{\label{sec:system}The system}

Our system of interest consists of $N$ organic molecules at positions $\mathbf{r}_n$, coupled to cavity modes with the in-plane momenta $\mathbf{k}_\Vert$. We consider the weak-pumping and weak system-environment coupling regimes. That is, the system is assumed to host at most one exciton at a time, and the exciton is weakly coupled to both the local phonon bath and the electromagnetic free-space modes outside the cavity. For simplicity, we also assume homogeneous transition dipole moments (TDMs). Taking both the singlets ($S$) and triplets ($T$) into account, we can describe the system with the Holstein-Tavis-Cummings (HTC) Hamiltonian $H=H_S+H_B+H_I$~\cite{Herrera2017,Takahashi2019}. Using the rotating-wave approximation and omitting the triplet-cavity mode couplings, we have
\begin{equation}
\begin{split}
H_S&=\sum_{n=1}^NE_s|S_n\rangle\langle S_n|+\sum_{n=1}^NE_t|T_n\rangle\langle T_n|+\sum_{\mathbf{k}_\Vert}E_c(\mathbf{k}_\Vert)\hat{a}_{\mathbf{k}_\Vert}^\dagger\hat{a}_{\mathbf{k}_\Vert}\\
&\hspace{10pt}+V_{st}\sum_{n=1}^N\big(|S_n\rangle\langle T_n|+|T_n\rangle\langle S_n|\big)\\
&\hspace{10pt}+\sum_{n,\mathbf{k}_\Vert}\Big(g(\mathbf{k}_\Vert)e^{i\mathbf{k}_\Vert\cdot\mathbf{r}_n}|S_n\rangle\langle\mathcal{G}|\hat{a}_{\mathbf{k}_\Vert}+\text{h.c.}\Big),
\label{eq:Hsystem}
\end{split}
\end{equation}
\begin{equation}
H_B=\sum_{n,l}\epsilon_{n,l}\hat{b}^\dagger_{n,l}\hat{b}_{n,l}+\sum_\mathbf{k}E_f(\mathbf{k})\hat{c}^\dagger_\mathbf{k}\hat{c}_\mathbf{k},\hspace{70pt}
\label{eq:Hbath}
\end{equation}
\begin{equation}
\begin{split}
H_I&=\sum_{n,l}\Big(\big(\sigma_{n,l}^s|S_n\rangle+\sigma_{n,l}^t|T_n\rangle\big)\langle\mathcal{G}|\hat{b}_{n,l}+\text{h.c.}\Big)\\
&\hspace{10pt}+\sum_{n,l}\big(\tilde{\sigma}_{n,l}^s|S_n\rangle\langle S_n|+\tilde{\sigma}_{n,l}^t|T_n\rangle\langle T_n|\big)(\hat{b}_{n,l}+\hat{b}^\dagger_{n,l})\\
&\hspace{10pt}+\sum_{n,\mathbf{k}}\Big(f(\mathbf{k})e^{i\mathbf{k}\cdot\mathbf{r}_n}|S_n\rangle\langle\mathcal{G}|\hat{c}_\mathbf{k}+\text{h.c.}\Big)
\label{eq:Hinteraction}
\end{split}
\end{equation}
Here, $E_s$ and $E_t$ are the \raisebox{0.01ex}{$^{*}$}0$\leftrightarrow$0 transition energies. This simplification makes it more straightforward and meaningful to compare the different coupling regimes. Namely, while it is typically some higher-order \raisebox{0.01ex}{$^{*}$}0$\to\xi$ (0$\to\xi^*$) transition that dominates weak (strong) coupling, focusing on the \raisebox{0.01ex}{$^{*}$}0$\leftrightarrow$0 transition allows us to consider both simultaneously within a unified framework and decide which one is better. 

$\hat{a}^\dagger_{\mathbf{k}_\Vert}$ ($\hat{a}_{\mathbf{k}_\Vert}$) is the creation (annihilation) operator of a photon with the energy $E_c(\mathbf{k}_\Vert)$, for which we have
\begin{equation}
E_c(\mathbf{k}_\Vert)=\frac{\hbar c}{n_\textit{eff}}\sqrt{\frac{m^2\pi^2}{L_c^2}+|\mathbf{k}_\Vert|^2}.
\label{eq:cav-energy}
\end{equation}
$\hbar$ is the reduced Planck’s constant, $c$ the speed of light
in vacuum, $n_\textit{eff}$ the refractive index of the emitting layer (which we assume constant), $L_c$ the cavity thickness, $m\in\mathbb{N}$, and the in-plane momentum is related to the outcoupling angle $\theta$ via
\begin{equation}
|\mathbf{k}_\Vert|=\frac{m\pi}{L_c}|\tan\theta|.
\label{eq:in-plane}
\end{equation}
For simplicity, we restrict our attention to the smallest possible energy, $m=1$.

$V_{st}$ is the singlet-triplet coupling strength. The global electronic ground state is denoted by $|\mathcal{G}\rangle$. Note that using the \textit{global} ground state prevents us from going beyond the single-excitation subspace.
\begin{equation}
g(\mathbf{k}_\Vert)=\mu\sqrt{\frac{E_c(\mathbf{k}_\Vert)}{2\epsilon_0V(\mathbf{k}_\Vert)}}
\label{eq:gN}
\end{equation}
is the light-matter coupling strength, with $\mu$, $\epsilon_0$, and $V({\mathbf{k}_\Vert})$ being the TDM, vacuum permittivity, and mode volume, respectively. The TDM of triplets is typically negligible~\cite{Bhuyan2023}, which allowed us to omit the triplet-cavity mode interactions. The mode volume can be evaluated as~\cite{Dutta2021}
\begin{equation}
    V(\mathbf{k}_\Vert)=\frac{\int\epsilon\langle\hat{E}(\mathbf{k}_\Vert)^2\rangle d^3\mathbf{r}}{\max\{\epsilon\langle\hat{E}(\mathbf{k}_\Vert)^2\rangle\}},
\end{equation}
where $\epsilon=n_\textit{eff}^2$ is the dielectric function and $\hat{E}(\mathbf{k}_\Vert)=\sqrt{E_c(\mathbf{k}_\Vert)/\big(\epsilon_0V(\mathbf{k}_\Vert)\big)}(\hat{a}_{\mathbf{k}_\Vert}+\hat{a}^\dagger_{\mathbf{k}_\Vert})\sin(\pi z/L_c)$ is the electric-field operator inside the cavity~\cite{Gerry2005}. $z$ is the distance from the cathode. Since we have assumed constant $n_\textit{eff}$, we find that $V({\mathbf{k}_\Vert})=V=AL_c/2$. Here, $A$ is the mode volume's effective cross-section.

$\hat{b}^\dagger_{n,l}$ ($\hat{b}_{n,l}$) is the creation (annihilation) operator of a localized phonon with the energy $\epsilon_{n,l}$. Similarly, $\hat{c}^\dagger_\mathbf{k}$ ($\hat{c}_\mathbf{k}$) is the creation (annihilation) operator of a free-space photon with the energy $E_f(\mathbf{k})=\hbar c|\mathbf{k}|$. $\sigma_{n,l}^{s(t)}$ and $\tilde{\sigma}_{n,l}^{s(t)}$ denote the coupling strengths associated with relaxation and pure dephasing, respectively, between the $l$th harmonic mode and a singlet (triplet) exciton at the molecular site $n$. Finally,
\begin{equation}
f(\mathbf{k})=\mu\sqrt{\frac{E_f(\mathbf{k})}{2\epsilon_0V_f}}    
\end{equation}
is the coupling strength of singlet excitons and free-space photons, with $V_f$ being a free-space quantization volume that gets canceled later on.


\subsection*{Detuning-corrected coupling strength}

Since one of our objectives is to assist with actual device design, the effects of cavity thickness should be taken more accurately into account. By tuning $L_c$, one can adjust $E_c(\mathbf{k}_\Vert)$ out of resonance with $E_s$, in which case the coupling strength should gradually vanish. However, this does not occur in Eq.~\eqref{eq:gN} as is.

Writing $H_S$ in the interaction picture and performing time coarse graining over some adequate timescale $\Delta t$~\cite{Majenz2013}, we get
\begin{equation}
    H_S'\approx\frac{1}{\Delta t}\sum_{n,\mathbf{k}_\Vert}\int_0^{\Delta t}\Big[g(\mathbf{k}_\Vert)e^{i(E_s-E_c(\mathbf{k}_\Vert))s/\hbar+i\mathbf{k}_\Vert\cdot\mathbf{r}_n}|S_n\rangle\langle\mathcal{G}|\hat{a}_{\mathbf{k}_\Vert}+\text{h.c.}\Big]ds.
\end{equation}
Assuming small disorder in the singlet energies, the detunings become independent. And because $N\gg0$, central limit theorem allows us to replace the uniform distributions with Gaussians. Evaluating the integrals and returning to the Schrödinger picture, we get
\begin{equation}
    \tilde{g}(\mathbf{k}_\Vert)=g(\mathbf{k}_\Vert)\exp\Bigg[{-\frac{1}{2}\Bigg(\frac{E_s-E_c(\mathbf{k}_\Vert)}{E_\textit{cut}(\mathbf{k}_\Vert)}\Bigg)^2}\Bigg].
    \label{eq:Gaussian}
\end{equation}
Here, $E_\textit{cut}(\mathbf{k}_\Vert)$ is a cut-off energy that should satisfy $E_\textit{cut}(\mathbf{k}_\Vert)\gg g(\mathbf{k}_\Vert)$ for the interaction-picture state to remain nearly constant over the averaging interval $\Delta t$~\cite{Wang2015}. Accordingly, we use $E_\textit{cut}(\mathbf{k}_\Vert)=1000g(\mathbf{k}_\Vert)$. While the in-depth analysis of $E_\textit{cut}(\mathbf{k}_\Vert)$ falls outside the scope of this article, this specific value---with the rest of the parameters---allows for a tuning range of a few tens of nanometers in cavity thickness, which is consistent with prior works (see, e.g., Refs.~\cite{Murawski2014,Mischok2023}). In actual experiments, $E_\textit{cut}(\mathbf{k}_\Vert)$ could be treated as a fitting parameter. In fact, Eq.~\eqref{eq:Gaussian} has been shown to provide good experimental fits in similar physical systems~\cite{Sang2021,Kratochwil2021}.

\subsection*{Diagonalizing the Hamiltonian in different coupling regimes}

Due to the weak system-environment couplings, $H_I$ can be treated perturbatively and $H_S$ diagonalized (nearly) independently of it. Assuming also weak singlet-triplet but strong cavity coupling, we get the following $N+1$ eigenstates in the singlet-cavity mode subspace,
\begin{align}
    |P_+(\mathbf{k}_\Vert)\rangle&=\frac{\alpha(\mathbf{k}_\Vert)}{\sqrt{N}}\sum_{n=1}^Ne^{i\mathbf{k}_\Vert\cdot\mathbf{r}_n}|S_n\rangle\otimes|0\rangle+\beta(\mathbf{k}_\Vert)|\mathcal{G}\rangle\otimes|1_{\mathbf{k}_\Vert}\rangle,
    \label{eq:UP}\\
    |P_-(\mathbf{k}_\Vert)\rangle&=\frac{\beta(\mathbf{k}_\Vert)}{\sqrt{N}}\sum_{n=1}^Ne^{i\mathbf{k}_\Vert\cdot\mathbf{r}_n}|S_n\rangle\otimes|0\rangle-\alpha(\mathbf{k}_\Vert)|\mathcal{G}\rangle\otimes|1_{\mathbf{k}_\Vert}\rangle,
    \label{eq:LP}\\
    |D_k(\mathbf{k}_\Vert)\rangle&=\frac{1}{\sqrt{N}}\sum_{n=1}^Ne^{i(2\pi nk/N+\mathbf{k}_\Vert\cdot\mathbf{r}_n)}|S_n\rangle\otimes|0\rangle.
    \label{eq:dark}
\end{align}
$|P_+(\mathbf{k}_\Vert)\rangle$ is the upper polariton, $|P_-(\mathbf{k}_\Vert)\rangle$ is the lower polariton, and $|D_k(\mathbf{k}_\Vert)\rangle$---with $k\in[1,N-1]$---are the dark states, which are collectively referred to as the exciton reservoir. The parameters $\alpha(\mathbf{k}_\Vert)$ and $\beta(\mathbf{k}_\Vert)$ satisfy
\begin{equation}
    |\alpha(\mathbf{k}_\Vert)|^2=1-|\beta(\mathbf{k}_\Vert)|^2=\frac{1}{2}\Bigg(1+\frac{E_s-E_c(\mathbf{k}_\Vert)}{\sqrt{\big(E_s-E_c(\mathbf{k}_\Vert)\big)^2+4\tilde{g}_N(\mathbf{k}_\Vert)^2}}\Bigg),
\end{equation}
with $\tilde{g}_N(\mathbf{k}_\Vert)=\sqrt{N}\tilde{g}(\mathbf{k}_\Vert)$, and the polariton eigenenergies are
\begin{equation}
    E_\pm(\mathbf{k}_\Vert)=\frac{E_s+E_c(\mathbf{k}_\Vert)}{2}\pm\sqrt{\tilde{g}_N(\mathbf{k}_\Vert)^2+\frac{\big(E_s-E_c(\mathbf{k}_\Vert)\big)^2}{4}}.
\end{equation}
The $N-1$ dark states, in turn, share the eigenenergy $E_s$. In the triplet subspace, we get the $N$ trivial eigenstates $|T_n\rangle$ with the eigenenergy $E_t$.

More formally, the system is said to be in the strong-coupling regime if
\begin{equation}
2\tilde{g}_N(\mathbf{k}_\Vert)>\frac{\hbar}{2}\big(\gamma+\kappa(\mathbf{k}_\Vert)\big),
\label{eq:boundary}
\end{equation}
where $\gamma$ and $\kappa(\mathbf{k}_\Vert)$ are the excitonic and photonic linewidths, respectively. We will return to both later in this article. The \textit{ultrastrong} coupling regime, which we do not consider in this article, is reached when $2\tilde{g}_N(\mathbf{k}_\Vert)>\min\{E_s,E_c(\mathbf{k}_\Vert)\}/5$~\cite{Bhuyan2023}. Here, the rotating-wave approximation would fail and our model would no longer be valid.

When $\tilde{g}_N(\mathbf{k}_\Vert)$ is too small to satisfy Eq.~\eqref{eq:boundary} (e.g., there are too few molecules or too large detuning) we enter the weak-coupling regime and treat also the cavity coupling perturbatively. While the dark states and triplets remain intact, we do the following replacements for the polaritons. If $E_s>E_c(\mathbf{k}_\Vert)$:
\begin{equation}
\begin{cases}
    |\alpha(\mathbf{k}_\Vert)|^2\to1,\\
    |P_+(\mathbf{k}_\Vert)\rangle\to|D_N(\mathbf{k}_\Vert)\rangle,\\
    |P_-(\mathbf{k}_\Vert)\rangle\to|\mathcal{G}\rangle\otimes|1_{\mathbf{k}_\Vert}\rangle.
\end{cases}
\end{equation}
If $E_s\leq E_c(\mathbf{k}_\Vert)$:
\begin{equation}
\begin{cases}
    |\alpha(\mathbf{k}_\Vert)|^2\to0,\\
    |P_+(\mathbf{k}_\Vert)\rangle\to|\mathcal{G}\rangle\otimes|1_{\mathbf{k}_\Vert}\rangle,\\
    |P_-(\mathbf{k}_\Vert)\rangle\to|D_N(\mathbf{k}_\Vert)\rangle.
\end{cases}
\label{eq:WC-transformation}
\end{equation}
That is, the sign of detuning dictates which polariton becomes the fully symmetric momentum state with $k=N$ (which, as we shall see, is not dark) and which the purely photonic Fock state.

Note that the weak-coupling picture remains valid even in the absence of cavity coupling ($g(\mathbf{k}_\Vert)=0$). In this case, there are simply no transitions between the excitons and the virtual ``cavity modes.'' It is also important to notice that, in reality, transitioning between the weak- and strong-coupling regimes (or the perturbative and hybridized regimes) should be treated in a more continuous fashion. While we leave this task for future studies, we hypothesize that a suitable $E_\textit{cut}(\mathbf{k}_\Vert)$ could provide a smooth bridge between the two regimes.

\subsection*{\label{sec:rates}Dynamics of open quantum systems}

The time evolution of open quantum systems $\rho(t)$, caused by inevitable interactions with the environment, is described in the weak-coupling, memoryless limit by the Gorini–Kossakowski–Sudarshan–Lindblad (GKSL) master equation~\cite{Petru2007}
\begin{equation}
    \frac{d}{dt}\rho(t)=-\frac{i}{\hbar}[H_S,\rho(t)]+\sum_j\Gamma_j\Big(\hat{L}_j\rho(t)\hat{L}_j^\dagger-\frac{1}{2}\{\hat{L}_j^\dagger\hat{L}_j,\rho(t)\}\Big).
    \label{eq:ME}
\end{equation}
The commutator $[H_S,\rho(t)]:=H_S\rho(t)-\rho(t)H_S$ gives the unitary dynamics of the system, while the sum over jump operators $\hat{L}_j$ gives the non-unitary, environment-induced dynamics. Each channel is weighted by the rate $\Gamma_j$, and the anti-commutator is defined as $\{X,Y\}:=XY+YX$. In our case, $\rho(t)$ describes the joint state of singlets, triplets, and cavity modes.

Under weak system-environment coupling, the rates of jump operators $|f\rangle\langle i|$ mapping eigenstates of $H_S$ to each other ($|i\rangle\mapsto|f\rangle$) can be calculated using Fermi's golden rule (FGR)~\cite{Fermi1950,Forrest2020},
\begin{equation}
\Gamma_{i\to f}=\frac{2\pi}{\hbar}|\langle f|H_I|i\rangle|^2\rho(E_\textit{if}).
\label{eq:FGR}
\end{equation}
Here, $\rho(E_\textit{if})$ is the joint density of states of the initial and final wavefunctions, not to be confused with $\rho(t)$.

\begin{figure}[t!]
\includegraphics[width=\linewidth]{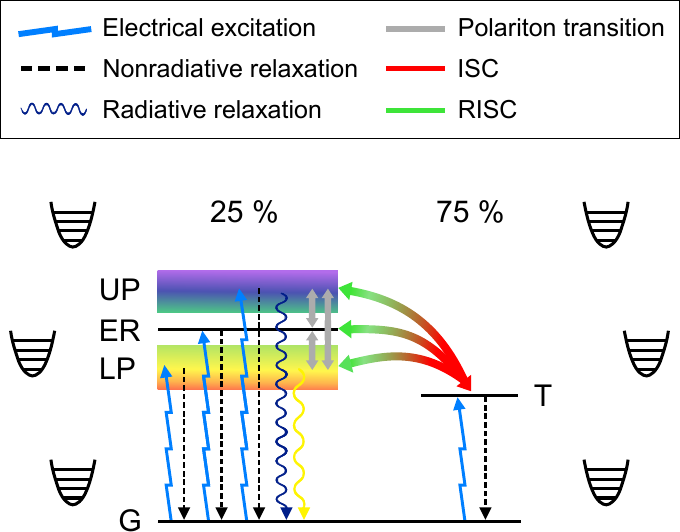}
\caption{\label{fig:processes}
Jablonski diagram of the system of interest: an organic molecule inside an optical cavity, embedded in a phonon bath and experiencing electrical excitation, polariton transitions, inter-system crossing (ISC), reverse inter-system crossing (RISC), emission, and non-radiative losses. Although a single, strongly coupled molecule is shown, we consider an ensemble of $N$ molecules across all the coupling regimes: no coupling, weak coupling, and strong coupling. It is important to note that the polaritons are collective states of all the $N$ sites and not localized, as depicted here for simplicity. UP = upper polariton, ER = exciton reservoir, LP = lower polariton, T = triplet state, G = ground state.
}
\end{figure}

Eqs.~\eqref{eq:ME} and~\eqref{eq:FGR} are the main tools of this article. We shall next derive all the different rates using (mainly) FGR. Then, substituting them into the GKSL master equation, we are able to solve the relevant population dynamics and estimate OLED performance. Fig.~\ref{fig:processes} illustrates (almost) all the processes we are interested in. The only process we ignore is polariton dephasing, but it is shown in the Supplementary information how this does not affect our results.

\subsection*{Electrical excitation}

Before the FGR calculations, let us derive simple rates for electrical excitation. Applying a current density $J$ to the mode volume's effective cross-section $A$ and assuming that all the injected electrons and holes recombine to excitons, $JA/e$ gives the total rate of exciton formation. Taking both the spin statistics and molecular sites into account, the pumping rate of singlets at a single molecular site becomes $\Gamma_{\mathcal{G}\otimes0\to S_n}=JA/(4eN)$. The corresponding jump operator is $|S_n\rangle\langle\mathcal{G}|$. Triplets, on the other hand, are incoherently pumped by $|T_n\rangle\langle\mathcal{G}|$ with trice as high rates, $\Gamma_{\mathcal{G}\otimes0\to T_n}=3JA/(4eN)$.

Next, projecting the local pumping rates to polariton basis, we get
\begin{equation}
\begin{split}
    \Gamma_{\mathcal{G}\otimes0\to X(\mathbf{k}_\Vert)}&=\sum_{n=1}^N|\langle X(\mathbf{k}_\Vert)|S_n\rangle|^2\Gamma_{\mathcal{G}\otimes0\to S_n}=|h_{X(\mathbf{k}_\Vert)}|^2\Gamma_{\mathcal{G}\otimes0\to S_n},\\
    X&=P_\pm,D_k.
\end{split}
\label{eq:pumping}
\end{equation}
$|h_{X(\mathbf{k}_\Vert)}|^2$ is the excitonic weight of the target state, e.g., $|\alpha(\mathbf{k}_\Vert)|^2$ for the upper polariton. Since the injected electrons and holes do not directly interact with the cavity modes, the in-plane momentum $\mathbf{k}_\Vert$ in Eq.~\eqref{eq:pumping} can be treated as effectively random.


\subsection*{Phonon-mediated transitions}

We consider two types of vibrational transitions: 1) polariton transitions arising from singlet dephasing and 2) non-radiative relaxation to the global ground state. In thermal equilibrium, the rates of transitions involving energy exchange read
(see Supplementary note 1)
\begin{equation}
\begin{split}
\Gamma_{X(\mathbf{k}_\Vert)\to Y(\mathbf{k}_\Vert)}&=\frac{|h_{X(\mathbf{k}_\Vert)}|^2|h_{Y(\mathbf{k}_\Vert)}|^2}{N}\tilde{\mathcal{J}}_{s}(|\Delta E|)\Big[\langle n(|\Delta E|)\rangle+\Phi(-\Delta E)\Big],\\
X,Y&=P_\pm,D_k.
\end{split}
\label{eq:phonon-trans}
\end{equation}
$|h_{X(\mathbf{k}_\Vert)}|^2$ and $|h_{Y(\mathbf{k}_\Vert)}|^2$ are, respectively, the excitonic weights of the initial and final state [cf. Eq.~\eqref{eq:pumping}], $\Delta E=E_{Y(\mathbf{k}_\Vert)}-E_{X(\mathbf{k}_\Vert)}$, $\tilde{\mathcal{J}}_s(\bullet)$ is the spectral density giving the strength of phonon coupling, $\langle n(\bullet)\rangle$ is the mean number of phonons stimulating the transition, and $\Phi(\bullet)$ is the Heaviside step function giving the spontaneous contribution for downward transitions. When $\Delta E=0$, the transition rate (or the rate of ``$k$-mixing'') is given by the sum of upward and downward rates.

Since dephasing arises from small-energy fluctuations due to acoustic phonons, we use the Ohmic spectral density
\begin{equation}
    \tilde{\mathcal{J}}_s(E)=\frac{10E}{\hbar}e^{-E/\tilde{E}_\textit{s,cut}},
\end{equation}
where $\tilde{E}_\textit{s,cut}$ is a cut-off energy. The mean phonon number in thermal equilibrium is given by the Bose-Einstein distribution
\begin{equation}
\langle n(E)\rangle=\frac{1}{e^\frac{E}{k_BT}-1},
\label{eq:BE}
\end{equation}
with $k_B$ and $T$ being the Boltzmann constant and temperature, respectively.

The FGR rates of nonradiative relaxation, on the other hand, read
\begin{equation}
\begin{split}
\Gamma_{X(\mathbf{k}_\Vert)\to\mathcal{G}\otimes0}&=|h_{X(\mathbf{k}_\Vert)}|^2\mathcal{J}_{s(t)}\big(E_{X(\mathbf{k}_\Vert)}\big)\Big[\big\langle n\big(E_{X(\mathbf{k}_\Vert)}\big)\big\rangle+1\Big],\\
X&=P_\pm,D_k,T_n.
\end{split}
\label{eq:phonon-relax}
\end{equation}
As the dissipative channels are provided by higher-energy optical phonons, we use the super-Ohmic spectral density
\begin{equation}
\mathcal{J}_{s(t)}(E)=\frac{E^3}{\hbar E_{{s(t)},\textit{cut}}^2}e^{-E/E_{{s(t)},\textit{cut}}},
\label{eq:superohmic}
\end{equation}
where $E_{{s(t),}\textit{cut}}$ denotes the cut-off energy. This form has been shown to agree well with experimental findings (see, e.g., Ref.~\cite{Leppala2024}).

\subsection*{ISC and RISC}

Even though the singlet-triplet interactions appear in $H_s$, FGR can be applied to calculate the rates of ISC and RISC as well. For simplicity, we take a semiclassical approach and describe the density of final states with the Gaussian disorder model (see our earlier work~\cite{Siltanen2025} for details). The resulting ISC rates can be expressed as
\begin{equation}
\begin{split}
    \Gamma_{X(\mathbf{k}_\Vert)\to T_n}=&\frac{|h_{X(\mathbf{k}_\Vert)}|^2}{N}\frac{V_{st}^2}{\hbar}\sqrt{\frac{\pi}{\lambda_{X(\mathbf{k}_\Vert)} k_BT}}\\
    &\times\exp\Bigg[-\frac{(\lambda_{X(\mathbf{k}_\Vert)}+E_t-E_{X(\mathbf{k}_\Vert)})^2}{4\lambda_{X(\mathbf{k}_\Vert)}k_BT}\Bigg],\hspace{5pt}X=P_\pm,D_k,
\end{split}
\label{eq:ISC}
\end{equation}
where $|h_{X(\mathbf{k}_\Vert)}|^2$ is, again, the excitonic weight of the participating state in the polariton branch and $\lambda_{X(\mathbf{k}_\Vert)}$ is the reorganization energy---the energy required by the system to reshape its nuclear configuration after the transition, independent from the transition's direction. In RISC, the sign of the energy difference $E_t-E_{X(\mathbf{k}_\Vert)}$ is just flipped. Note that these rates are essentially the Marcus electron transfer rates~\cite{Marcus1956} multiplied by the excitonic weights and diluted by the collective nature of strong coupling.

In Ref.~\cite{Eizner2019}, the authors defined the polaritonic reorganization energies as $\big(\sqrt{\lambda_{s}}+\sqrt{\lambda_{g}}\big)^2$, where $g$ stands for the electronic ground state. This definition, however, does not take into account the different excitonic/photonic contents of the upper and lower polaritons. For example, either one can be fully excitonic, in which case the reorganization energy should just be $\lambda_{s}$. Hence, we define the polaritonic reorganization energies differently.

The reorganization energy is more generally defined as~\cite{Forrest2020}
\begin{equation}
    \lambda_{i,f}=\frac{1}{2}\zeta(Q_i-Q_f)^2,
\end{equation}
where $\zeta$ is the curvature of the potential energy surfaces (same for initial and final states) and $Q_{i(f)}$ is the nuclear coordinate of the initial (final) state. Motivated by the polaritonic composition, we model the upper polaritons' nuclear coordinates as the convex combination $Q_+(\mathbf{k}_\Vert)=|\alpha(\mathbf{k}_\Vert)|^2Q_s+|\beta(\mathbf{k}_\Vert)|^2Q_g$ and similarly for the lower polaritons. Using these coordinates, it is quite straightforward to show that
\begin{align}
\lambda_{P_+(\mathbf{k}_\Vert)}&=\big(|\alpha(\mathbf{k}_\Vert)|^2\sqrt{\lambda_s}+|\beta(\mathbf{k}_\Vert)|^2\sqrt{\lambda_g}\big)^2,\\
\lambda_{P_-(\mathbf{k}_\Vert)}&=\big(|\beta(\mathbf{k}_\Vert)|^2\sqrt{\lambda_s}+|\alpha(\mathbf{k}_\Vert)|^2\sqrt{\lambda_g}\big)^2.
\end{align}

\subsection*{Optical transitions}

We consider three types of optical transitions: 1) the singlet excitons emitting directly to free space, 2) the singlet excitons emitting to the cavity mode, and 3) a photon in the cavity mode leaking out of the cavity. Importantly, the second transition only occurs in the weak-coupling regime. Furthermore, as there is no degree of freedom mediating these transitions, the in-plane momentum remains constant in all cases.

The FGR calculations (see Supplementary note 2) reveal that only the symmetric momentum states $|D_N(\mathbf{k}_\Vert)\rangle$ can emit (both to free space and cavity mode). The rest remain dark due to destructive interference. The jump operators in the free-space case are $|\mathcal{G}\rangle\otimes|0\rangle\langle P_\pm(\mathbf{k}_\Vert)|$, while the corresponding rates read
\begin{equation}
\Gamma_\textit{free}^\pm(\mathbf{k}_\Vert)=N|h_\pm(\mathbf{k}_\Vert)|^2\frac{\mu^2E_\pm(\mathbf{k}_\Vert)^3}{\pi\epsilon_0\hbar^4c^3}.
\end{equation}
$|h_\pm(\mathbf{k}_\Vert)|^2$ is the emitting polariton's excitonic weight. In a more complete model, the lower polariton might overlap with the free-space emission spectrum, enabling the lower polariton's radiative pumping~\cite{Perez2025}. Here, however, there is no such overlap and we can omit photon recycling.

In the cavity case, the jump operator becomes $|\mathcal{G}\rangle\otimes|1_{\mathbf{k}_\Vert}\rangle\langle D_N(\mathbf{k}_\Vert)|$ and
\begin{equation}
\Gamma_\textit{cavity}(\mathbf{k}_\Vert)=NF_P(\mathbf{k}_\Vert)\frac{\mu^2E_c(\mathbf{k}_\Vert)^3}{\pi\epsilon_0\hbar^4c^3}\exp\Bigg[{-\Bigg(\frac{E_s-E_c(\mathbf{k}_\Vert)}{E_\textit{cut}(\mathbf{k}_\Vert)}\Bigg)^2}\Bigg].
\label{eq:weak-rate}
\end{equation}
Here, $F_P(\mathbf{k}_\Vert)$ is the Purcell factor~\cite{Gerry2005,Barrett2020}
\begin{equation}
    F_P(\mathbf{k}_\Vert)=\frac{3}{4\pi^2}\frac{Q(\mathbf{k}_\Vert)}{V}\Big(\frac{2L_c\cos\theta}{m}\Big)^3,
\label{eq:Purcell}
\end{equation}
$m=1$, and $Q(\mathbf{k}_\Vert)$ is the cavity quality factor. Assuming symmetric mirrors with the reflectivity $R<1$, we have~\cite{Forrest2020}
\begin{equation}
    Q(\mathbf{k}_\Vert)=\frac{E_c(\mathbf{k}_\Vert)L_c}{\hbar c}\frac{\sqrt{R}}{1-R}.
\end{equation}
Even if we treat the TDMs as fixed, the singlets couple to both s- and p-polarized cavity modes, and no net preference exists at ensemble level. For this reason, we use the mirror reflectivity~\cite{Hecht2017}
\begin{equation}
R=\frac{1}{2}\Bigg(\Bigg|\frac{n_\textit{eff}\cos\theta-n_\textit{mirr}\cos\phi}{n_\textit{eff}\cos\theta+n_\textit{mirr}\cos\phi}\Bigg|^2+\Bigg|\frac{n_\textit{mirr}\cos\theta-n_\textit{eff}\cos\phi}{n_\textit{mirr}\cos\theta+n_\textit{eff}\cos\phi}\Bigg|^2\Bigg),
\label{eq:Fresnel}
\end{equation}
where $n_\textit{mirr}$ and $\phi$ are the refractive index of the mirrors and the angle of transmission, respectively. When $R=0$, there is no mechanism populating the cavity modes and we are in the ``no-coupling regime,'' i.e., we have a basic OLED (cf. Fig.~\ref{fig:schematics}).

In both coupling regimes, photon leakage is described by the annihilation operator $\hat{a}_{\mathbf{k}_\Vert}$ and the rate~\cite{Gu2021}
\begin{equation}
\kappa_\pm(\mathbf{k}_\Vert)=|h_\mp(\mathbf{k}_\Vert)|^2\frac{c}{2L_c}\frac{1-R}{\sqrt{R}}.
\label{eq:kappa}
\end{equation}
Note that here we have used the \textit{photonic} weight of the emitting polariton. Note also that here we have not considered where the leaked photon eventually ends up. It may escape the entire device, contributing to external quantum efficiency (EQE), get reabsorbed~\footnote{If the radiative loss rate of photons is much larger than the total decoherence rate of singlets, we are in the ``bad-cavity limit'' and reabsorption can be safely ignored~\cite{Grange2015}.}, or couple to waveguide modes or surface plasmon polaritons (SPPs). In this work, however, we are interested in the \textit{internal} quantum efficiency (IQE). In fact, conventional microcavity-based strategies aimed at enhancing IQE often induce additional absorption and mode leakage (guided or evanescent), which reduces outcoupling efficiency and thus EQE~\cite{Manish2025}.

\subsection*{Excitonic and photonic linewidths}

We now have almost all the ingredients needed to evaluate $\rho(t)$ and IQE. To determine which coupling regime the system is in, we still need to evaluate the excitonic and photonic linewidths more rigorously [cf. Eq.~\eqref{eq:boundary}]. While $\kappa(\mathbf{k}_\Vert)$ is directly given by Eq.~\eqref{eq:kappa}, it follows from the standard relation between longitudinal and transverse relaxation times~\cite{Francis1986,Grange2015,Grandi2016} that
\begin{equation}
\gamma=\frac{1}{2}(\Gamma_r+\Gamma_\textit{s,nr}+\Gamma_\textit{ISC})+\Phi,
\end{equation}
where the first three rates are, respectively, the rates of radiative relaxation, non-radiative relaxation, and inter-system crossing of uncoupled singlet excitons. $\Phi$---or the inverse of transverse relaxation time---is the rate of pure dephasing, and we estimate it in Supplementary note 3. In particular, we derive the decoherence function $\kappa_{n}(t)=\langle\mathcal{G}|\rho(t)|S_n\rangle/\langle\mathcal{G}|\rho(0)|S_n\rangle$ and observe it to be Gaussian. We therefore fit $\exp(-\Phi^2t^2)$ to it, which yields $\Phi$.

\subsection*{Open-system populations and IQE}

The open-system populations can be solved by substituting all the jump operators with their corresponding rates to the GKSL master equation~\eqref{eq:ME} and sandwiching appropriately. Ignoring unitary dynamics and dephasing (see Supplementary note 4 for justification), the resulting system of coupled rate equations can be compactly expressed as
\begin{equation}
\begin {split}
\frac{d}{dt}\langle X(\mathbf{k}_\Vert,t)\rangle&=\langle X(\mathbf{k}_\Vert)|\frac{d}{dt}\rho(t)|X(\mathbf{k}_\Vert)\rangle\\
&\approx\sum_{Y\neq X}\Big[\Gamma_{Y(\mathbf{k}_\Vert)\to X(\mathbf{k}_\Vert)}\langle Y(\mathbf{k}_\Vert,t)\rangle-\Gamma_{X(\mathbf{k}_\Vert)\to Y(\mathbf{k}_\Vert)}\langle X(\mathbf{k}_\Vert,t)\rangle\Big],\\
X&=P_\pm,D_k,T_n,\mathcal{G}\otimes0.\label{eq:rate-eqs-1}
\end{split}
\end{equation}
Under steady-state conditions, the rate equations become even simpler,
\begin{equation}
\sum_{Y\neq X}\Gamma_{Y(\mathbf{k}_\Vert)\to X(\mathbf{k}_\Vert)}\langle Y(\mathbf{k}_\Vert)\rangle=\sum_{Y\neq X}\Gamma_{X(\mathbf{k}_\Vert)\to Y(\mathbf{k}_\Vert)}\langle X(\mathbf{k}_\Vert)\rangle.
\label{eq:rate-eqs-2}
\end{equation}
Together with the normalization condition $\sum_X\langle X(\mathbf{k}_\Vert)\rangle=1$, these equations have a unique solution.

\begin{figure*}[t!]
\centering
\includegraphics[width=\linewidth]{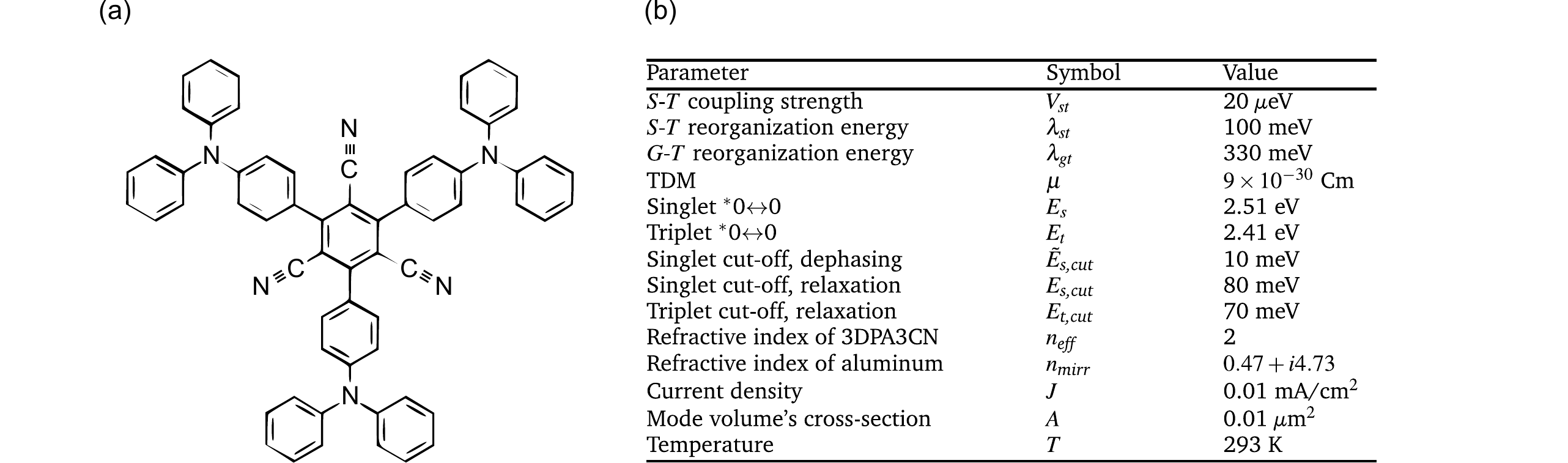}
\caption{\label{fig:specs}
(a) Chemical structure of 3DPA3CN. (b) Parameters used in this article.
}
\end{figure*}

Solving for the steady-state populations and assuming both a uniform distribution of generated in-plane momenta and that they remain constant until relaxation\footnote{$\mathbf{k}_\Vert$ could theoretically change in ISC-RISC cycles, but this should not affect the IQE, as we average in the end.}, the IQE---i.e., the ratio of generated photons to injected electrons---becomes the arithmetic average
\begin{equation}
\text{IQE}=\frac{1}{K}\sum_{\lambda=\pm}\sum_{\mathbf{k}_\Vert}\frac{\big(\Gamma_\textit{free}^\lambda(\mathbf{k}_\Vert)+\kappa_\lambda(\mathbf{k}_\Vert)\big)\langle P_\lambda(\mathbf{k}_\Vert)\rangle}{JA/e}\times100\%.
\label{eq:SC-IQE}
\end{equation}
Here, $K$ is the number of ``momentum bins.'' Although given in terms of polariton populations, it is important to notice how this formulation of IQE behaves in the weak- and no-coupling regimes. Going first to weak coupling and then to no coupling, we get
\begin{align}
\text{IQE}&\to\frac{1}{K}\sum_{\mathbf{k}_\Vert}\frac{\kappa_+(\mathbf{k}_\Vert)\langle\mathcal{G}\otimes1_{\mathbf{k}_\Vert}\rangle+\Gamma_\textit{free}^-(\mathbf{k}_\Vert)\langle D_N(\mathbf{k}_\Vert)\rangle}{JA/e}\times100\%\\
&\to\frac{1}{K}\sum_{\mathbf{k}_\Vert}\frac{\Gamma_\textit{free}^-(\mathbf{k}_\Vert)\langle D_N(\mathbf{k}_\Vert)\rangle}{JA/e}\times100\%=\frac{\Gamma_r\langle S\rangle}{JA/e}\times100\%.
\end{align}
That is, the hybrid polaritonic contributions first split into purely excitonic and photonic parts, whereas the final sum is solely excitonic. Here we assumed positive detuning ($E_c(\mathbf{k}_\Vert)>E_s\hspace{3pt}\forall\mathbf{k}_\Vert$) just for the sake of simpler notation.

\subsection*{Simulation results}

We apply our master equation model for the example molecule 1,3,5-tris(4-(diphenylamino)phenyl)-2,4,6-tricyanobenzene [3DPA3CN, see Fig.~\ref{fig:specs}(a)] due to its thorough characterization in the existing literature~\cite{Taneda2015,Eizner2019}. As for the mirrors, we use aluminum. The material-specific parameters are listed in Fig.~\ref{fig:specs}(b), along with other example parameters. We consider the range of $L_c$ from 100 nm to 150 nm; $E_s=E_c(0)$ at 123.49 nm. We also consider the range of $N$ from $5\times10^4$ to $5\times10^5$; with more molecules, we would enter the ultrastrong coupling regime. 

\begin{figure}[t!]
\includegraphics[width=\linewidth]{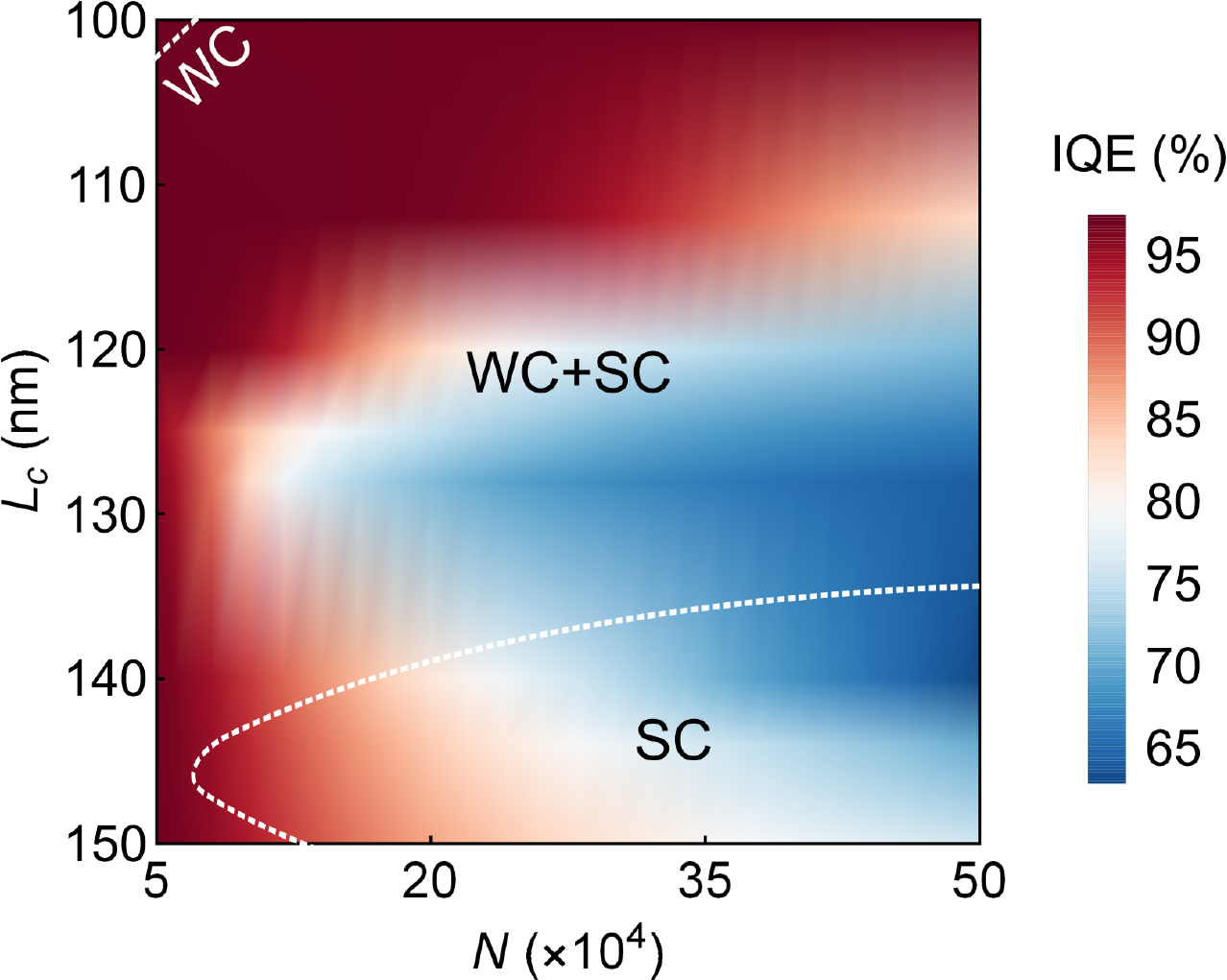}
\caption{\label{fig:iqe-contour}
IQE as a function of cavity thickness $L_c$ and the number of coupled molecules $N$. The white dashed curves separate regions, where the system is entirely in the weak-coupling regime (WC), strong-coupling regime (SC), or different regimes at different outcoupling angles (WC+SC).
}
\end{figure}

\begin{figure*}[t!]
\includegraphics[width=\linewidth]{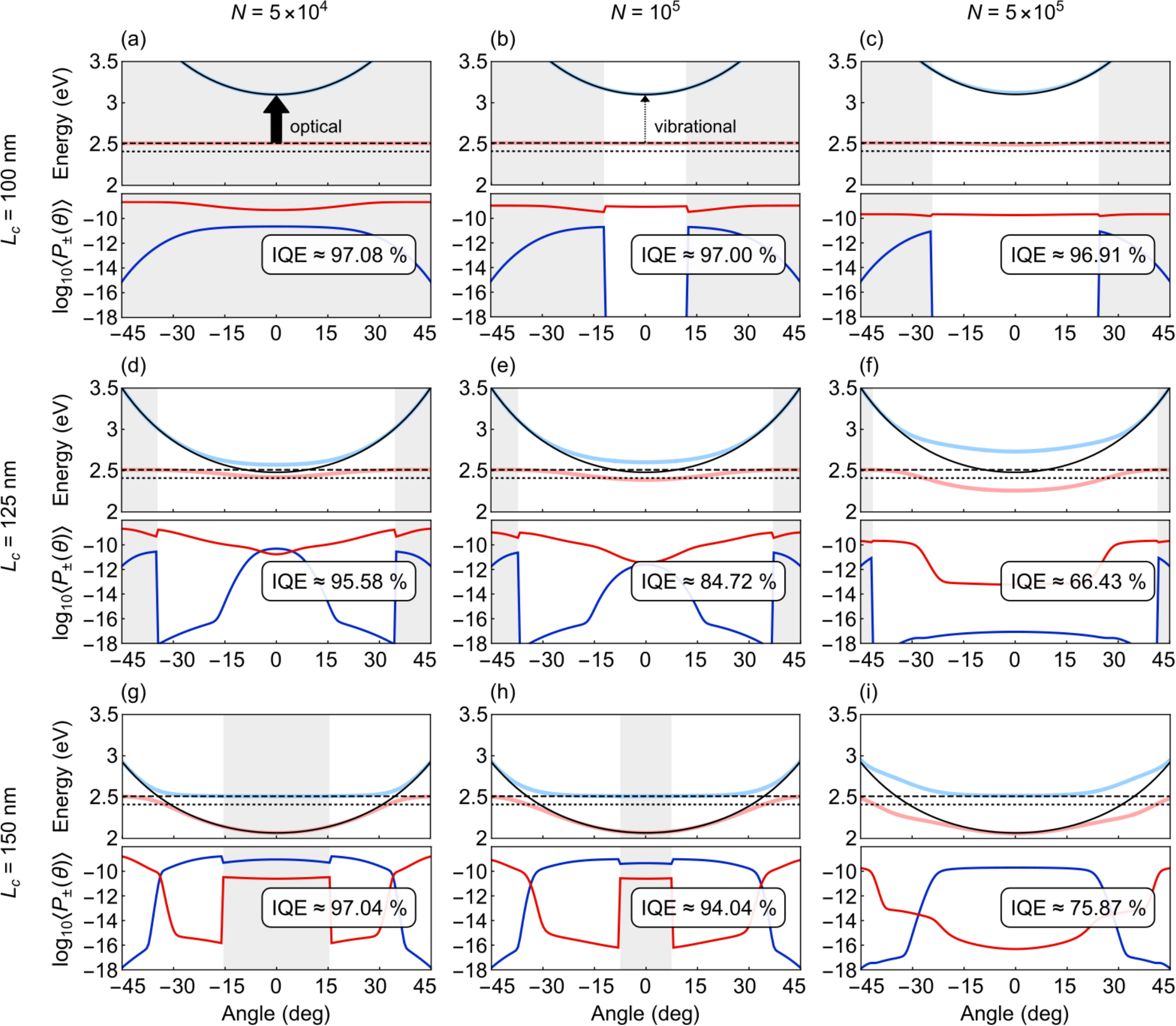}
\caption{\label{fig:iqes}
Polariton energies and steady-state populations as functions of the outcoupling angle $\theta$, shown for different numbers of coupled molecules $N$ and cavity thicknesses $L_c$. The insets also show the angle-averaged IQEs, where $K=313$. Light blue = $E_+(\theta)$, light red = $E_-(\theta)$, solid black = $E_c(\theta)$, dashed black = $E_s$, dotted black = $E_t$, dark blue = $\langle P_+(\theta)\rangle$, dark red = $\langle P_-(\theta)\rangle$, gray shading = weak-coupling regime, no shading = strong-coupling regime.
}
\end{figure*}

The simulation results are presented in Figs.~\ref{fig:iqe-contour} and ~\ref{fig:iqes} for $K=313$. The color map in Fig.~\ref{fig:iqe-contour} shows $\text{IQE}(L_c,N)$, with the white dashed curves separating regions where the system is entirely in the weak-coupling regime (WC), strong-coupling regime (SC), or a combination of both (WC+SC). That is, the boundary condition~\eqref{eq:boundary} holds for some angles but not all. Strictly speaking, the system can never be \textit{fully} in the strong-coupling regime, as the detuning explodes with $|\theta|\to90^\circ$. Hence, we say that the system is ``entirely'' in the strong-coupling regime if the boundary condition~\eqref{eq:boundary} holds for $|\theta|\leq45^\circ$.

Fig.~\ref{fig:iqe-contour} provides a clear answer to the research question: \textit{Does stronger light-matter coupling translate into better device performance?} Namely, the collective coupling strength can be increased by raising $N$ or reducing $L_c$. With too small $L_c$, though, the large detuning may suppress the coupling. Regardless of the method, it is evident from the SC regime in Fig.~\ref{fig:iqe-contour} that stronger coupling strengths mean \textit{lower} IQEs---even lower than the reference IQE of 96.87~\% with no cavity coupling (see Supplementary note 5). In contrast, the maximum IQE of 97.41~\% is achieved in the WC regime (also calculated in Supplementary note 5).

Although the rates of emission in both the WC and SC regimes are comparable, they are weighted by the emitting states' populations. Hence, it is the rates at which these states become populated \textit{themselves} that explains the discrepancy between the WC and SC IQEs. Transitioning from $|D_N\rangle$ to $|\mathcal{G}\rangle\otimes|1_{\mathbf{k}_\Vert}\rangle$ in the WC regime occurs much more efficiently than from the exciton reservoir to the polaritons in the SC regime, especially with large Rabi splittings [cf. Eqs.~\eqref{eq:phonon-trans} and~\eqref{eq:weak-rate}]. And even if the latter occurred efficiently, it would mean that the polariton is mostly excitonic, i.e., its overall emission rate would lack the fast $\kappa_\pm(\mathbf{\mathbf{k}_\Vert})$ component [cf. Eq.~\eqref{eq:kappa}]. The situation, however, might become exactly the opposite with small enough $E_\textit{cut}(\mathbf{k}_\Vert)$ and large enough $\tilde{E}_\textit{s,cut}$. With suitable cut-off energies, the vibrational transition rates might outperform their Purcell-enhanced, optical counterparts. In addition, one could enhance the polaritonic RISC rates with small $N$ (see also Ref.~\cite{Siltanen2025}).

Fig.~\ref{fig:iqes} shows the eigenenergies, their steady-state populations, and the angle-averaged IQEs for nine different pairs of $L_c$ and $N$, matching with Fig.~\ref{fig:iqe-contour}. The WC regimes are highlighted with gray shading.

Fig.~\ref{fig:iqes}, especially panels (a) and (b), illustrates well the previous reasoning. As soon as an interval of strong coupling opens---though described in this work as a sudden transition---the transition rate to the emitting upper polariton collapses. And while the transition rate to the lower polariton slightly increases, the state is mostly excitonic. Hence, the overall photon production rate and IQE cannot increase. In fact, they slightly decrease. In all cases, the steady-state populations and therefore IQEs are well explained by both the coupling regime and the distance of the energy levels from $E_s$ [cf. Eqs.~\eqref{eq:phonon-trans} and~\eqref{eq:weak-rate}].

We would like to highlight another interesting detail in panels (g) and (h). In both cases, two SC regimes close to the $E_s=E_c(\mathbf{k}_\Vert)$ resonance are separated by a WC regime, where the detuning becomes too large to sustain strong coupling. That is, energy anti-crossing does not necessarily indicate a uniform SC regime.

\section*{Conclusions}

In this work, we introduced the first unified quantum master equation model for microcavity OLEDs in different light-matter coupling regimes. Specifically, we derived the rates for electrical excitation, polariton transitions, non-radiative losses, ISC, RISC, and emission, assuming weak pumping and weak system-environment coupling. We solved the population dynamics by incorporating these rates, along with the corresponding jump operators, into the GKSL master equation. We applied our model to calculate and compare the IQE of 3DPA3CN in all the coupling regimes. Restricting to the \raisebox{0.01ex}{$^{*}$}0$\leftrightarrow$0 transition---common for weak and strong coupling---allowed us to decide which coupling regime is the best. As an interesting side result, we found that energy anti-crossing, typical hallmark of strong coupling, does not necessarily indicate uniform strong coupling.

With conventional parameter choices, the highest IQE was achieved in the weak-coupling regime. It might be surprising at first, that the worst IQE was achieved in the strong-coupling regime. Yet it makes perfect sense, as the polaritons should first be populated, and---in the absence of photon recycling/radiative pumping---that can be very inefficient due to low-energy acoustic phonons. However, as suggested, which coupling regime performs best depends strongly on the used parameters. With suitable cut-off energies, for example, polaritons in the strong-coupling regime could be populated more efficiently than bare cavity modes in the weak-coupling regime. Furthermore, as $N=1$ maximizes the rate of RISC directly to the lower polariton, distributed single-molecule strong coupling in organic optoelectronics would represent a fascinating direction of follow-up research.

Expanding the model is as important as it is challenging. While we leave this task for future studies, here we speculate on how such a model could be constructed. 1) Here we assumed the recombination efficiency of 100 \%. Relaxing this assumption, the IQEs simulated in this article should actually be interpreted as the exciton-to-photon conversion efficiencies (not electron-to-photon). More realistic recombination efficiencies could be estimated, e.g., with drift diffusion models~\cite{Rossi2020} or kinetic Monte Carlo simulations~\cite{Taherpour2024}. 2) Here we also assumed aligned TDMs. Taking inhomogeneous coupling strength and thermal disorder into account, the dark states would become brighter~\cite{Houdre1996,Khazanov2023,Dutta2024}. 3) A more realistic model should also include all the relevant transition energies and internal conversions. With sufficiently weak phonon couplings, transition rates between these energy levels could still be estimated with FGR. Stronger phonon couplings, as well as the intermediate cavity coupling regime omitted in this article, could be treated, e.g., with non-Markovian quantum state diffusion~\cite{Leppala2024}. 4) Achieving higher luminances---and eventually efficiency roll-off---would require stronger pumping rates and moving beyond the linear regime. Most notably, annihilation processes involving singlets, triplets, and polarons would become critical~\cite{Murawski2013}. However, diagonalizing the HTC Hamiltonian in the strong-coupling regime quickly becomes a formidable task as the number of excitations grows, necessitating the use of permutation symmetries~\cite{two-exc-eigen}, mean-field approximations~\cite{Hu2020}, or hierarchical equations of motion~\cite{Tanimura2020}. 5) As the IQE represents the upper bound of EQE---an end-user-relevant quantity---future theoretical investigations should also address losses to waiveguides and SPPs and, eventually, outcoupling efficiency. Our model could be combined, e.g., with transfer matrix methods to explore this aspect~\cite{Krummacher2009}. 6) It would be very interesting to see how ultrastrong coupling would influence the photodynamics and device performance~\cite{Mischok2023}.


Nonetheless, it is precisely due to its simplicity that our unified model provides a robust framework for understanding and optimizing various types of microcavity OLEDs, with considerable potential to guide the design of more efficient light-emitting devices.

\section*{Author contributions}

O.S. and K.S.D. conceived the work. O.S. performed the theoretical analysis. K.L. oversaw the theoretical analysis. O.S. wrote the article with input from K.S.D. K.S.D. supervised the work. All authors discussed the results and contents of the article.

\section*{Conflicts of interest}

There are no conflicts to declare.

\section*{Data availability}

Data sharing is not applicable to this article as no datasets were generated or analyzed during the current study.

\section*{Acknowledgements}

O.S. acknowledges fruitful discussions with T. Lepp\"{a}l\"{a}, A. Dutta, and H. Lyyra. This project was funded by the European Research Council through the European Union’s Horizon 2020 research and innovation program (Grant Agreement Number 948260) and partially by the European Innovation Council through the SCOLED project (Grant Agreement Number 101098813). Views and opinions expressed are, however, those of the authors only and do not necessarily reflect those of the European Union or the European Innovation Council. Neither the European Union nor the granting authority can be held responsible for them.







\bibliography{references} 
\bibliographystyle{rsc} 

\end{document}